\documentclass[12pt]{iopart}
\usepackage{iopams}
\usepackage{epsfig}   
\usepackage{graphics}
\begin{document}

\title[Supernova limits on brane world cosmology]{Supernova limits on brane world cosmology}

\author{Malcolm Fairbairn and Ariel Goobar}

\address{Department of Physics, Stockholm University, Albanova
         University Center \\
         S--106 91 Stockholm, Sweden}

\begin{abstract}
By combining the first year data from the Supernova Legacy Survey (SN LS) 
and the
recent detection of the baryon acoustic peak in the Sloan Digital Sky Survey
data, we are able to place strong constraints on models where the cosmic
acceleration is due to the leakage of gravity from the brane into the bulk
on large scales.  In particular, we are able to show that the DGP model is
not compatible with a spatially flat universe.  We generalize our
analysis to phenomenological toy models where the curvature of the brane
enters into the Friedmann equations in different ways.
\end{abstract}

\eads{\mailto{malc@physto.se}, \mailto{ariel@physto.se}}




\section{Introduction}
Observations of Type Ia supernovae have led us to understand that the
universe started to accelerate at a redshift of $z\sim 1$
\cite{riess98,perl99,knop03,riess04}.  
This suggests the presence of some energy
component which redshifts slowly and which had the same density as
matter relatively recently, $z<1$.  The desire to explain this
behaviour has led to the development of a number of models such as
quintessence where the energy density corresponds to the potential
energy of some scalar field.

At the same time, models with large extra dimensions into which only
gravity can propagate have become popular which suggests that there may
be short distance corrections to gravity \cite{add,rs}.  When the
Friedmann equations are solved in these higher dimensional backgrounds,
the expression for the Hubble parameter changes to $H^2= G(\rho+k
\rho^2)$ where $k$ is related to the tension of the brane upon which
the standard model particles reside \cite{brane}.  However,
modifications of this form do not give rise to the kind of behaviour
required to explain the observed acceleration.

A related theory is that due to Dvali, Gabadadze and Porrati (DGP),
which presents a class of brane theories where the theory remains four
dimensional at short distances but crosses over to a higher
dimensional theory for gravity at some very large distance \cite{dgp}.
In this model, gravity leaks out into the bulk only at large distances
and the resulting modifications to the Friedmann equations can give
rise to universes with acceleration \cite{def2,def3}.

In section \ref{dgp1} we will briefly present the DGP model and then
we will show in section \ref{obs1} that is it marginally disfavoured
by observations of supernovae and baryon oscillations, at least in the
case of a spatially flat universe.  In section \ref{dgp2} we will try
to generalise the 1st (tt) Friedmann equation obtained in the DGP model to see
if the data indicates what kinds of modifications to the Friedmann
equations might explain the data.

\section{The DGP model\label{dgp1}}

The DGP theory starts with the idea that standard model gauge fields
are confined to live on a (3+1)D brane residing in an infinite (4+1)D
bulk, with different scales of gravity on the brane and in the bulk
\cite{dgp}.  The gravitational part of the action is therefore given
by
\begin{equation}
S=\frac{1}{2\mu^2}\int d^5z\sqrt{-g_5} R_5+\frac{1}{2\kappa^2}\int
d^4x\sqrt{-g_4}R_4
\end{equation}
where $\mu^2=M_{Pl}^{-2}$ and $\kappa^2=M_5^{-3}$.  The Friedmann
equation which comes from this gravitational action coupled to matter
on the brane take the form \cite{def1}
\begin{equation}
\epsilon\sqrt{H^2+\frac{k}{a^2}}=\frac{\kappa^2}{2\mu^2}\left(H^2+\frac{k}{a^2}\right)-\frac{\kappa^2}{6}\rho
\label{dgpfried}
\end{equation}
where $\epsilon=\pm 1$.  For our purposes it is important to note at
this point that the corrections to the normal equations involve the
extrinsic curvature of the brane itself, in other words $H^2+k
a^{-2}$. Later when we try and generalise this equation to compare
with the data, this observation will loosely motivate the form of our
generalisation.  If we set $\epsilon=1$ we can re-arrange things so
that the Friedmann equation takes the form
\begin{equation}
H^2+\frac{k}{a^2}=\left(\sqrt{\frac{\rho}{3M_{Pl}^2}+\frac{1}{4r_c^2}}+\frac{1}{2r_c}\right)^2
\end{equation}
where the length scale $r_c$ is defined to be
\begin{equation}
\frac{1}{2r_c}=\frac{\mu^2}{\kappa^2}
\end{equation}
and $r_c$ sets a length beyond which gravity starts to leak out into
the bulk, modifying the Friedmann equations. The relationship between
the Hubble expansion parameter today $H_0$ and that at a redshift $z$
is then given by
\begin{equation}
\frac{H(z)^2}{H_0^2}=H'(z)^2=\Omega_k(1+z)^2+\left(\sqrt{\Omega_M(1+z)^3+\Omega_{r_c}}+\sqrt{\Omega_{r_c}}\right)^2
\label{eq:leak1}
\end{equation}
where we use the usual notation $\Omega_x=\rho_x/\rho_{crit}$ where
$\rho_x$ refers to the energy density in component $x$ today and $\rho_{crit}=3H_0^2/8\pi G$.  The new
unusual parameter $\Omega_{r_c}$ takes the form
\begin{equation}
\Omega_{r_c}=\frac{\mu^4}{\kappa^4 H_0^2}
\end{equation}
and the constraint equation between the various components of energy
density at $z=0$ is given by
\begin{equation}
\Omega_M+\Omega_k+2\sqrt{\Omega_{r_c}}\sqrt{1-\Omega_k}=1
\end{equation}
So that, in particular for a flat universe with $k=0$ we have the relation
\begin{equation}
\Omega_{r_c}=\frac{(1-\Omega_M)^2}{4}
\end{equation}
It has been shown that by setting the length scale $r_c$ close to the horizon size, this extra contribution to the Friedmann
equation leads to acceleration which can in principle explain the supernova data \cite{def2,def3}.  In the next section we will see if this theory can fit with the latest observations.

\section{Confronting the model with cosmological data\label{obs1}}

Measurements of the brightness of Type Ia supernovae as a function of 
redshift are sensitive to the cosmological model via the integration over expansion history in the expression for the luminosity distance ($c=1$)
\begin{equation} 
 d_L = 
   \frac{1+z}{H_0\sqrt{|\Omega_k|}} {\cal S}
    \left( \sqrt{|\Omega_k|} \int_{0}^{z} { {d\tilde{z} \over H'(\tilde{z})}} \right)   
\label{eq:dl}
\end{equation}
where the function ${\cal S}(x)$ is defined as $\sin(x)$ for 
$\Omega_k<0$ ({\em closed Universe}), 
 $\sinh(x)$ for $\Omega_k >0$ ({\em open Universe}) and  
${\cal S}(x) = x$, and the factor
 $\sqrt{|\Omega_k|}$ is removed for the {\em flat Universe}

\begin{figure}
\begin{center}
\resizebox{0.7\textwidth}{!}{\includegraphics{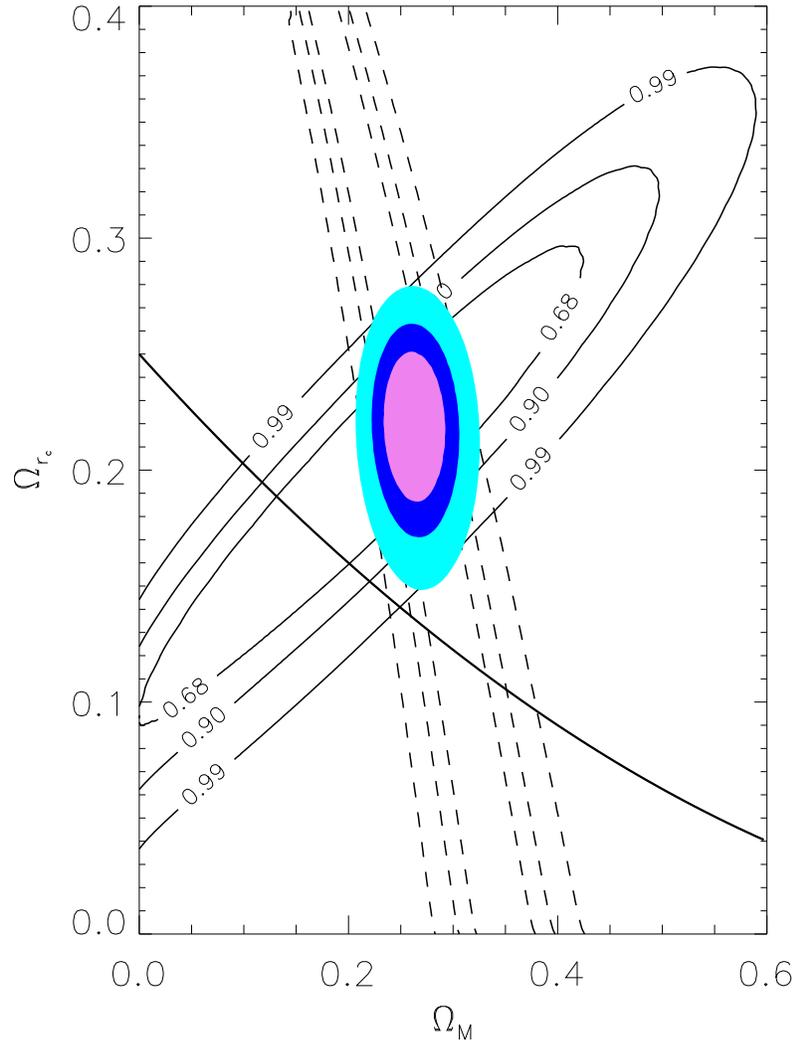}}

\caption{The solid curves show the 
allowed parameter regions in the $\Omega_M -\Omega_{r_c}$ plane
at the 68, 90 and 99\% confidence level. Solid thin contours correspond to the
the first year SNLS data in Astier et al (2005)\cite{astier05}, the dashed
lines show the corresponding regions from the baryon oscillation peak
in Eisenstein et al (2005) \cite{Eisenstein05}. The coloured contours
indicate the result of the combination of both data-sets. The thick solid
lines indicate the expected relation between $\Omega_M$ and $\Omega_{r_c}$
in a flat universe.}
\end{center}
\label{fig:alpha1}
\end{figure}

In the DGP model, $H(z)$ is then given by Eq.(\ref{eq:leak1}). Very
recently, the SNLS collaboration published the first ear data of its
planned five year survey (Astier et al, 2005) \cite{astier05}. The
data set includes 71 high redshift Type Ia supernovae (SNIa) in the
redshift range z=[0.2,1] and 44 low redshift SNIa compiled from the
literature but analysed in the same manner as the high-z
sample. Thanks to the multi-band, rolling search technique and careful
calibration, this data-set is arguably the best high-z SNIa
compilation to date, indicated by the very tight scatter around the
best fit in the Hubble diagram and the careful estimate of systematic
uncertainties in \cite{astier05}.  In our analysis (as in
\cite{astier05}) we fit for cosmological parameters with the prior
from the baryon oscillation peak detected in the SDSS Luminous Red Galaxy
survey (LRG) of Eisenstein el al (2005) \cite{Eisenstein05}, given by

\begin{equation}
{ \sqrt{\Omega_M} \over E(z_1)^{1 \over 3} }
\left[{1 \over z_1 \sqrt{|\Omega_k|} } 
{\cal S}\left(\sqrt{|\Omega_k|}\int_0^{z_1} {dz \over E(z)} \right) \right]^{2 \over 3} =0.469 \pm 0.017,
\label{eq:bao}
\end{equation}
where $E(z) = H(z)/H_0$ and $z_1=0.35$. Furthermore, ${\cal S}$ and $\Omega_k$ are defined
as in Eq.(\ref{eq:dl}). The quoted uncertainty corresponds to one
standard deviation, where a Gaussian probability distribution has been
assumed.
Figure \ref{fig:alpha1} shows the allowed regions in the 
$\Omega_{r_c}$--$\Omega_M$ plane in the DGP model from each of the two
data sets and the combined limits. Also shown in the graph is the 
curve corresponding to flatness, i.e. $\Omega_k=0$. 

The allowed region of the $\Omega_{r_c}$--$\Omega_M$ plane deduced from baryon oscillations forms a band whereas the allowed region from supernovae luminosity measurements forms an ellipsoid.  The fact that the baryon oscillations constrain models to lie on a band reflects the fact that they have only so far been detected at two redshifts, $z=0.35$ in the SDSS data and $z\sim 1100$ in the CMB data.  This gives us one ratio vs. the many ratios which come from the many type 1a supernovae.  However, the power of this approach lies in the fact that the band of allowed parameters from baryon oscillations is almost perpendicular to the major axis of the ellipsoid coming from the supernovae data.  This occurs in a very similar way in the conventional $\Omega_\Lambda$--$\Omega_M$ plane.  The use of the baryon oscillation prior therefore allows us to constrain model universes much more effectively.

\begin{figure}
\begin{center}
\resizebox{0.7\textwidth}{!}{\includegraphics{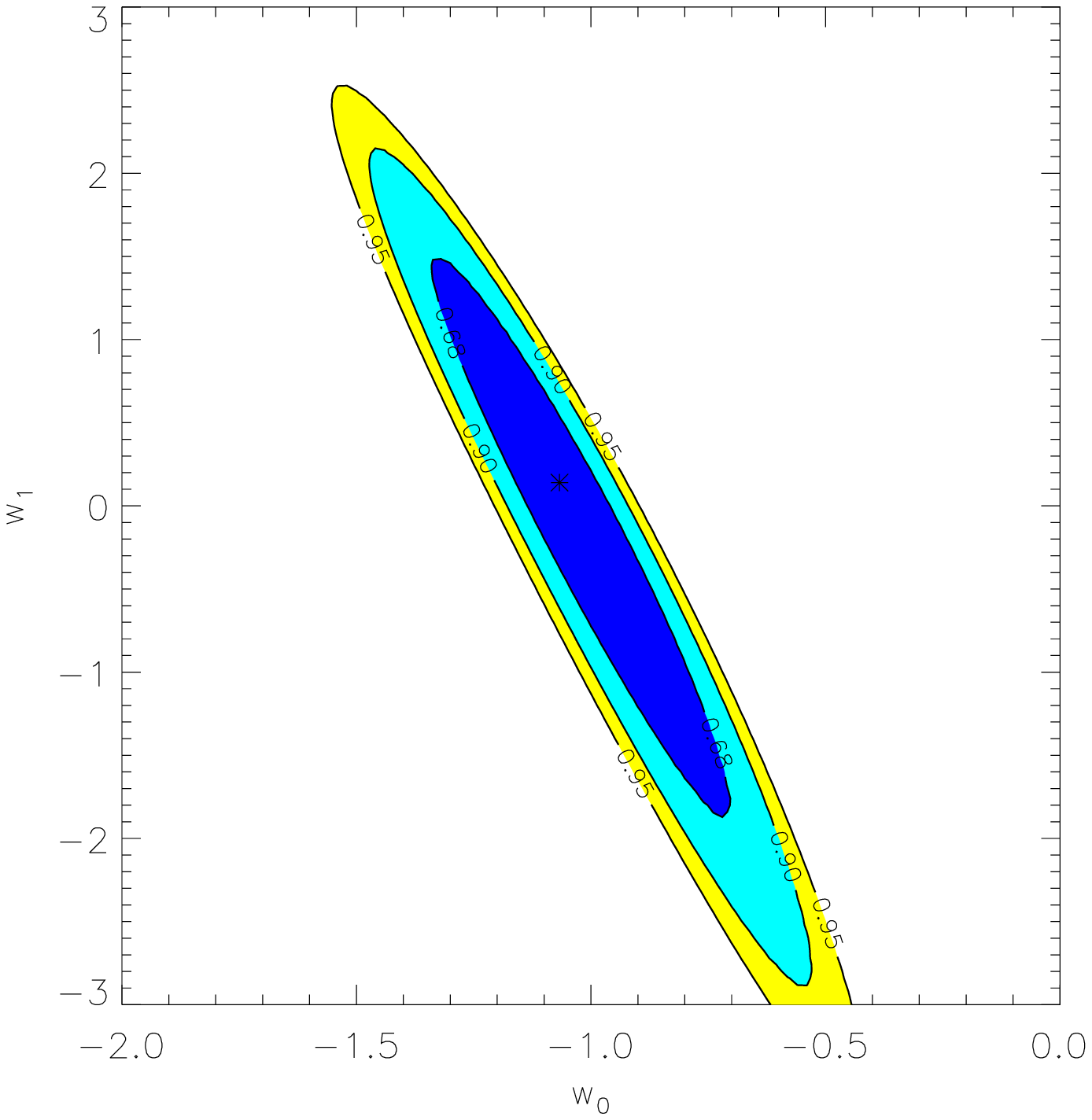}}
\caption{Allowed 68, 90 and 95 \% confidence regions in the $w_0$,$w_1$ 
parameter space for
the first year SNLS data with baryon oscillation prior in a flat universe. The
linear expansion $w(z) = w_0 + w_1\cdot z$ has been assumed. The best fit
solution $(w_0=-1.07,w_1=0.14)$ is also shown.}
\end{center}
\label{fig:w0w1}
\end{figure}

We conclude that
the fit of the DGP model to the supernova data along with a prior from the
baryon oscillation results is incompatible with a flat universe.  
Given the evidence for the latter from the CMB experiments, this model
is clearly disfavoured by the current observational data. For comparison,
in Figure\ref{fig:w0w1} we show the fit to the same data to a dark energy
model generically described by an equation of state parameter, 
$w(z) = w_0 + w_1\cdot z$,  where a flat universe is assumed. The $\Lambda$ model
$(w_0=-1, w_1=0)$ gives an excellent fit to the data. All the fits were 
performed with a modified version of the cosmology fitter (SNALYS) 
in the SNOC package \cite{snoc}.

\section{Generalisations of the DGP model\label{dgp2}}
 We would like to consider generalisations of the DGP model in order
to see if the data favours any particular modification of the Friedmann
equation.  (This is in the spirit of \cite{dvaliturner} and
\cite{chung}.)  The extra term in the Friedmann equation
(\ref{dgpfried}) comes from the derivative of the scale factor in the
5th dimension, which is set by the Israel branching conditions across
the brane and therefore by the extrinsic curvature on the brane. We
will therefore assume the extra term enters in the following form
\begin{equation}
\left(H^2+\frac{k}{a^2}\right)^{\alpha/2}=\frac{\kappa_*^2}{2\mu^2}\left(H^2+\frac{k}{a^2}\right)-\frac{\kappa_*^2}{6}\rho_m
\label{modfried}
\end{equation}
where now $\alpha$ is a free parameter and $\kappa_*^2$ is now related to some phenomenological mass scale $\tilde{M}$  
\begin{equation}
\kappa_*^2=\tilde{M}^{\alpha-4}.
\end{equation}
The expression for the Hubble expansion as a function of redshift is then
\begin{equation}
\frac{H(z)^2}{H_0^2}+\frac{k}{a^2H_0^2}-\frac{2\mu^2}{\kappa_*^2H_0^{2-\alpha}}\left(\frac{H(z)^2}{H_0^2}+\frac{k}{a^2H_0^2}\right)^{\alpha/2}=\Omega_M(1+z)^3
\end{equation}
or equivalently
\begin{equation}
\frac{H(z)^2}{H_0^2}-\Omega_k(1+z)^2-2\sqrt{\Omega_{r_c}}\left(\frac{H(z)^2}{H_0^2}-\Omega_k(1+z)^2\right)^{\alpha/2}=\Omega_M(1+z)^3
\label{eq:freealpha}
\end{equation}
where now we have generalised the expression for the DGP brane world dark energy density fraction 
\begin{equation}
\sqrt{\Omega_{r_c}}=\frac{\mu^2}{\kappa_*^2H_0^{2-\alpha}}
\end{equation}
which means that the constraint between the various densities at $z=0$ is given by
\begin{equation}
\Omega_M+\Omega_k+2\sqrt{\Omega_{r_c}}(1-\Omega_k)^{\alpha/2}=1
\end{equation}
In order to make sure there are no pathologies here, we need to ensure that
\begin{equation}
\Omega_k<1-\Omega_M\qquad ;\qquad (\Omega_M>0)
\end{equation}
then if we write
\begin{equation}
h=\sqrt{\frac{H(z)^2}{H_0^2}+\frac{k}{a^2H_0^2}}
\end{equation}
we can then solve the following equation as a function of redshift 
\begin{equation}
h^2- 2\sqrt{\Omega_{r_c}} h^{\alpha}=\Omega_M(1+z)^3
\end{equation}
and use the solution to obtain the luminosity distance.  Since $\alpha$ is no longer $\alpha=1$ we now obtain the solutions to this equation numerically.

\begin{figure}
\begin{center}
\resizebox{0.7\textwidth}{!}{\includegraphics{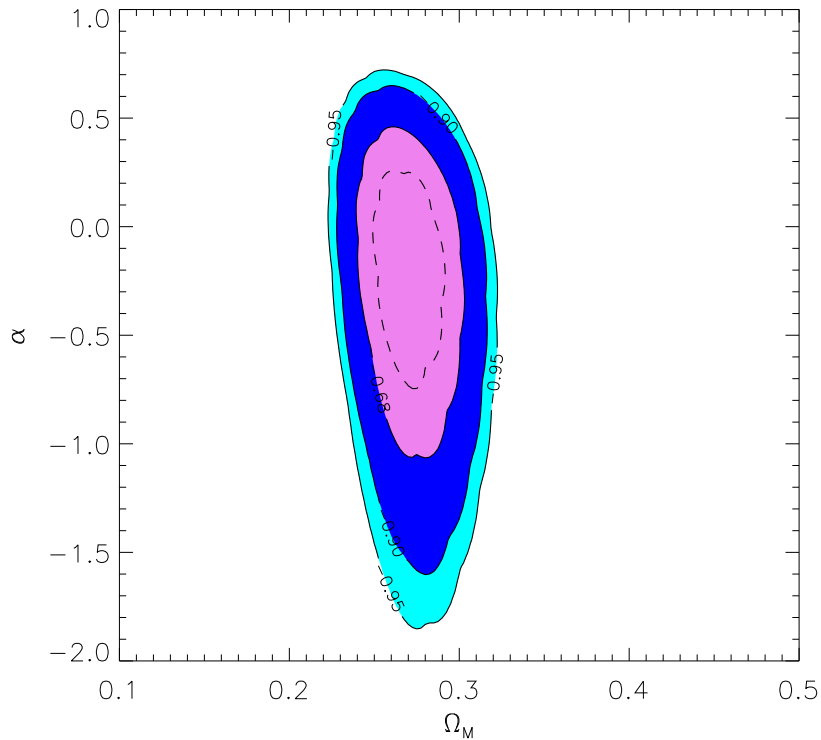}}
\caption{Allowed parameter regions in the $\alpha$--$\Omega_M$ plane
at the 1 $\sigma$ (one parameter), 68, 90 and 95\% confidence level for a flat universe.}
\end{center}
 \label{fig:alphafree}
\end{figure}

Figure \ref{fig:alphafree} shows the allowed values in the $\alpha$--$\Omega_M$
parameter space if a flat universe prior is imposed. The best fits are
centered close to $\alpha=0$, which corresponds to a constant in Eq.(\ref{eq:freealpha}), i.e. equivalent to $\Lambda$. The full data set of the SNLS, 
$\sim$700 spectroscopically confirmed  Type Ia SNe,
will soon provide further constraints in the possible values of the 
exponent $\alpha$.

\section{Conclusions}
The most recent high-redshift supernova data has been used to test the DGP
brane model of cosmology where dark energy is replaced by an extension of
the Friedmann equation to account for the leakage of gravity to a large
extra dimension. Using a prior from the SDSS baryon oscillation results 
we find that the this model is incompatible with the data for a flat universe.
Generalizations of the model including an extra term 
$H^\alpha/r_c^{2-\alpha}$ yields $-0.8 < \alpha < 0.3$ (1 $\sigma$) i.e 
consistent with $\alpha=0$, the situation corresponding to a non-redshifting cosmological constant.  

This is not completely surprising if we look at the generalised Friedmann
equation (\ref{modfried}) since during the matter dominated era, it is straightforward to show that the
equation of state $w=-1+{\alpha \over 2}$ \cite{dvaliturner}.  Observational
constraints on $w$ already push us close to $w \sim-1$ (see Figure \ref{fig:w0w1}) so we see that we require some small value of $\alpha$ in
order to match the observations at this time.

Finally, we note that it may be interesting in the future to see if it is possible to obtain a good fit to the data in models which give rise to different modifications to the Friedmann equations.

\ack
M.F. is grateful for funding from the Swedish Research Council (Vetenskapsr\aa det).  A.G. is a Royal Swedish Academy Research Fellow supported by a grant from the Knut and Alice
Wallenberg Foundation. We thank Rahman Amanullah for assistance in the
implementation of the DGP model in the SNALYS/SNOC cosmology fitter.  Also thanks to Shinji Tsujikawa for spotting typos in the first version.

\end{document}